\documentclass[a4paper,twocolumn,english,pre,nofootinbib]{revtex4}
\usepackage[T1]{fontenc}
\usepackage[latin9]{inputenc}
\usepackage{color}
\usepackage{babel}
\usepackage{mathtools}
\usepackage{amsmath}
\usepackage{amssymb}
\usepackage{graphicx}
\usepackage[unicode=false,pdfusetitle,
 bookmarks=false,bookmarksnumbered=false,bookmarksopen=false,
 breaklinks=false,pdfborder={0 0 1},backref=false,colorlinks=true]
 {hyperref}
\graphicspath{{./Figures/}}

\begin{document}


\title{Anomalous finite-size scaling in higher-order processes with absorbing states}

\author{Alessandro Vezzani}
\affiliation{Istituto dei Materiali per l'Elettronica ed il Magnetismo (IMEM-CNR), Parco Area delle Scienze, 37/A-43124 Parma, Italy}
\affiliation{Dipartimento di Scienze Matematiche, Fisiche e Informatiche,
Universit\`a degli Studi di Parma, Parco Area delle Scienze, 7/A 43124 Parma, Italy}
\affiliation{INFN, Gruppo Collegato di Parma, Parco Area delle Scienze 7/A, 43124 Parma, Italy}

\author{Miguel A. Mu{\~n}oz}
\affiliation{Departamento de
  Electromagnetismo y F{\'\i}sica de la Materia }
  \affiliation{Instituto Carlos I
  de F{\'\i}sica Te{\'o}rica y Computacional, Universidad de Granada.
  E-18071, Granada, Spain}

\author{Raffaella Burioni}
\affiliation{Dipartimento di Scienze Matematiche, Fisiche e Informatiche,
Universit\`a degli Studi di Parma, Parco Area delle Scienze, 7/A 43124 Parma, Italy}
\affiliation{INFN, Gruppo Collegato di Parma, Parco Area delle Scienze 7/A, 43124 Parma, Italy}

\date{\today}

\begin{abstract}
Here we study standard and higher-order birth-death processes on fully-connected networks, within the perspective of large-deviation theory (also referred
to as Wentzel-Kramers-Brillouin (WKB) method in some contexts). We obtain a general expression for the leading and next-to-leading terms of the stationary probability distribution of the fraction of "active" sites as a function of parameters and network size $N$.  We reproduce several results from the literature and, in particular, we derive all the moments of the stationary distribution for the $q$-susceptible-infected-susceptible ($q-SIS$) model, i.e., a high-order epidemic model requiring of $q$ active ("infected") sites to activate an additional one. We uncover a very rich scenario for the fluctuations of the fraction of active sites, with  non-trivial finite-size-scaling properties. In particular, we show that the variance-to-mean ratio
diverges at criticality for $[1 \leq q\leq 3]$, with a maximal variability at $q=2$, confirming that complex-contagion processes can exhibit peculiar scaling features including wild variability. Moreover, the leading-order in a large-deviation approach does not suffice to describe them: next-to-leading terms are essential to capture the intrinsic singularity at the origin  of systems with absorbing states. Some possible extensions of this work are also discussed.

\end{abstract}
\maketitle

\section{Introduction}

Systems with absorbing or quiescent states have played a central role
in the development of the theory of non-equilibrium
phase transitions \cite{Marro,Hinrichsen,GGMA,Henkel,Odor}.
Analysis of such systems is crucial to shed light onto apparently diverse phenomena such as catalytic reactions, the propagation of epidemics in complex networks, neural dynamics, viral spreading of memes in social networks, the emergence of consensus, desertification processes, and the transition to turbulence, to name but a few examples \cite{Liggett,Castellano,Vespi,Ziff,cardy1985,martinello,Romu,Radicchi,notarmuzi,juhasz,eluding,turbulence}. In particular, birth-death processes (or "creation-annihilation" particle processes)  on complex networks represent an extremely general and versatile framework to tackle such a variety of problems, as exemplified
by, e.g., models of epidemic propagation in which infected ("active") individuals can either heal (become "inactive") or infect their neighbors at some given rates, and all dynamics ceases in the absence of infection, i.e., once the absorbing or quiescent state has been reached.

The focus of attention in this context has recently shifted to the study of higher-order interactions (beyond simple pairwise ones) in the probabilistic rules for the birth-and-death processes; i.e. to include the possibility that more than one active site is required to generate further activations \cite{BATTISTON2020,BATTISTON2021}. Indeed, it has been shown that the presence of higher-order interactions (also called  "complex-contagion" processes \cite{centola2007,centola2010,Vespi,karsai,maxi,Mancastroppa1}) can lead to a change on the nature of the phase transition for a wide class of models describing, e.g., epidemics, opinion dynamics, synchronization, population-dynamics, etc.

For instance, the requirement of more than one single "active" (or "infected") individual needed to generate further activations (infections) gives typically rise to discontinuous or abrupt transitions with coexistence between quiescent and active states and hysteresis phenomena (see e.g. \cite{Henkel,Odor,Fiore,Jensen,Paula}).
Simplicial complexes and hypergraphs represent a natural and alternative framework to analyse these processes \cite{Bianconi16,BianconiDor20,hyper} with important
implications in research fields such as theoretical ecology \cite{Jacopo} and neuroscience \cite{Bassett}.

Theoretical analyses of these transitions often start from the consideration of complete or fully-connected graphs, for which the "ideal" mean-field dynamics is formally recovered in the limit of infinitely-large network sizes, $N$, allowing also to analyze finite-size corrections. Results for the dynamics of higher-order process on the complete graph have been obtained in recent years, but they are rather scattered in the literature.  Here, we recover many of these results by employing systematically a \emph{large-deviation} framework \cite{Touchette} (also called Wentzel-Kramers-Brillouin (WKB) method in  the context of e.g. population dynamics see, e.g., \cite{Kubo1973fluctuation,Assaf_2017,Dykman,McKane}) and study in detail several aspects of the most general birth-death processes, exhibiting a phase transition into an absorbing state.

In particular, we obtain a general expression for the leading and next-to-leading terms of the stationary probability distribution of the fraction of "active" sites, as a function of the systems size $N$.  By doing this, we first reproduce diverse results from the literature and, then, we also derive all the moments of the stationary distribution for the specific case of the $q$-susceptible-infected-susceptible ($q-SIS$) model, i.e., a higher-order epidemic model requiring of $q$ active ("infected") sites with $q>1$, to activate an additional one. We uncover a very rich phenomenology for the fluctuations of the fraction of positive sites, with a non-trivial dependence  both on the system size $N$ (i.e. anomalous finite-size scaling) and on the order $q$ of the interaction. In particular, we stress the fact that, crucially and contrarily to the standard situation, e.g. in equilibrium statistical mechanics, one needs to go beyond leading order in $N$ to properly describe critical fluctuations.

The paper is organized as follows. In Section \ref{Master}, we first introduce the general framework for a general birth-death process on a complete graph, deriving as a first step general results for the stationary distribution at large $N$ using a large-deviation approach \cite{Touchette}. In Section \ref{abs} we consider the case of systems  with an absorbing state {and we define a quasi-stationary distribution}. In Section \ref{Fluct} we study in detail the higher-order $q-SIS$ model, deriving all the moments of the { quasi-stationary} distribution and their finite-size scaling properties, underlining its non-trivial behavior. Finally, Section \ref{conc} summarizes the conclusions and some open problems.

\section{The Master Equation in the  large-deviation (or WKB) approach}
\label{Master}

In order to fix notation and ideas, let us recapitulate some well-known approaches and results \cite{Gardiner,vK,Romu,Kubo1973fluctuation,Assaf_2017,Dykman,Kamenev,McKane}.
For this, let us consider a dynamical process on a fully-connected network (or "complete graph") of size $N$. The network state is specified by 
  a set of binary variables $\sigma_i=0,1$: one for each node $i$. The variable $n=\sum_i \sigma_i$ counts the number of active nodes, i.e., in state $\sigma_i=1$.  
The transition-rate functions $\gamma^-(n)$ and $\gamma^+(n)$, represent the probability that $n$ decreases or increases by one unit, respectively, defining a general  mean-field-like dynamics on the complete graph, as determined by the (one-step) Master equation \cite{Gardiner,vK}:
\begin{eqnarray}
&& P(n,t+1)-P(n,t)= - P(n,t)(\gamma^+(n)+\gamma^-(n))
\nonumber \\ && + P(n+1,t)\gamma^-(n+1) + P(n-1,t)\gamma^+(n-1).
\label{eq:master}
\end{eqnarray}
for the probability to be in the state $n$ at time $t$, $P(n,t)$. Observe that Eq.\eqref{eq:master} may describe many possible mean-field-like models such as, e.g.,  the Ising model, the  SIS model, the voter model, and also models with more complex behavior involving higher-order interactions on $q$ sites, such as  the q-neighbor Ising model \cite{qIsing} or the q-voter model \cite{qVoter}. The associated stationary distribution, $P_{st}(n)$ 
is simply given by the detailed-balance condition \cite{Gardiner,vK}:
\begin{equation}
P_{st}(n) \gamma^+(n)=P_{st}(n+1)\gamma^-(n+1).
\label{eq:stat}
\end{equation}
with $\gamma^-(0)=\gamma^+(N)=0$, since $0\leq n \leq N$, an equation that can  be formally solved in an exact way:
\begin{equation}
P_{st}(n)= P_{st}(0) \prod_{j=1}^n \gamma^+(j-1)/\gamma^-(j),
\label{eq:prod}
\end{equation}
where $P_{st}(0)$ is fixed by the overall normalisation condition (note, in particular, that if $\gamma^+(n_0)=0$ for some $n_0$, this implies that  $P_{st}(n) =0$ for $n>n_0$ and, if $\gamma^-(n_0)=0$, then $P_{st}(n) =0$ for $n<n_0$, so that the dynamics is asymptotically confined in a subset of the state space).  Eq.\eqref{eq:prod} can be used to obtain an exact numerical evaluation of the stationary probability distribution; indeed, it has been employed in different contexts such as for the q-neighbor Ising model \cite{qIsing}, for the SIS model  and its generalizations \cite{SIS1prod,SIS2prod,SISVerhulst} and for neutral models in ecology \cite{Neutral},
to name but a few examples.

To make further progress, let us assume that, in the limit of large network sizes ($N \rightarrow \infty$),  $\gamma^+(n)$ and $\gamma^-(n)$ just depend on the fraction of active sites, $x=n/N$
(that can be treated as a continuous variable),
so that Eq.(\eqref{eq:stat}) can be written as
\begin{equation}
P_{st}(x,N)~ \gamma^+(x)=P_{st}(x+1/N,N) ~ \gamma^-(x+1/N).
\label{eq:stat2}
\end{equation}
Within a large-deviation or WKB approach, at large $N$ $P_{st}(x,N)$ can be expressed as  \cite{Touchette,Assaf_2017,Dykman}
\begin{equation}
    P_{st}(x,N) = e^{\displaystyle{-NF(x)-g(x) + \Theta(1/N)}}.
\end{equation}
Plugging this expression into Eq.\eqref{eq:stat2}, one readily obtains:
\begin{eqnarray}
&&\log(\gamma^+(x)) -N F(x) - g(x) = \nonumber \\ && \log(\gamma^-(x+\frac{1}{N}))  -N F(x+\frac{1}{N})- g(x+\frac{1}{N})
\label{eq:stat3}
\end{eqnarray}
and, expanding Eq.\eqref{eq:stat3} for large $N$:
\begin{eqnarray}
&& \log(\gamma^+(x)) - N F(x) - g(x) = \nonumber \\ && \log(\gamma^-(x)) + \frac{\dot{\gamma}^-(x)}{\gamma^-(x)}\frac{1}{N} - N F(x) - N \dot{F}(x)\frac{1}{N}- 
\nonumber \\ && \frac{N}{2} \ddot{F}(x)\frac{1}{N^2} - g(x)
-\dot{g}(x)\frac{1}{N}
\label{eq:stat4}
\end{eqnarray}
where the dot stands for $x$-derivatives. 
Finally, equating terms of the same order in $1/N$  and performing the integrals, 
leads to:
\begin{eqnarray}
{F}(x)&=&C+ \int_c^x \log(\gamma^-(x')) - \log(\gamma^+(x'))dx' \nonumber
\\ 
g(x)&=&B+\frac{1}{2} \log(\gamma^-(x)\gamma^+(x)) 
\label{eq:stat7}
\end{eqnarray}
where $B$, $C$ and $0<c<1$ are arbitrary constants, so that
the stationary distribution reads \cite{Assaf_2017}:
\begin{equation}
P_{st}(n) \approx P_{S}(x,N) = K \frac{e^{-N \int_c^x \displaystyle{\log(\gamma^-(x')/\gamma^+(x'))dx'}}}{\sqrt{\gamma^-(x)\gamma^+(x)}}
\label{eq:stat8}
\end{equation}
where the constant $K$ (that depends on $c$)  is determined by the normalization condition. 
\begin{figure}
	\centering{
	\includegraphics[width=0.49\textwidth]{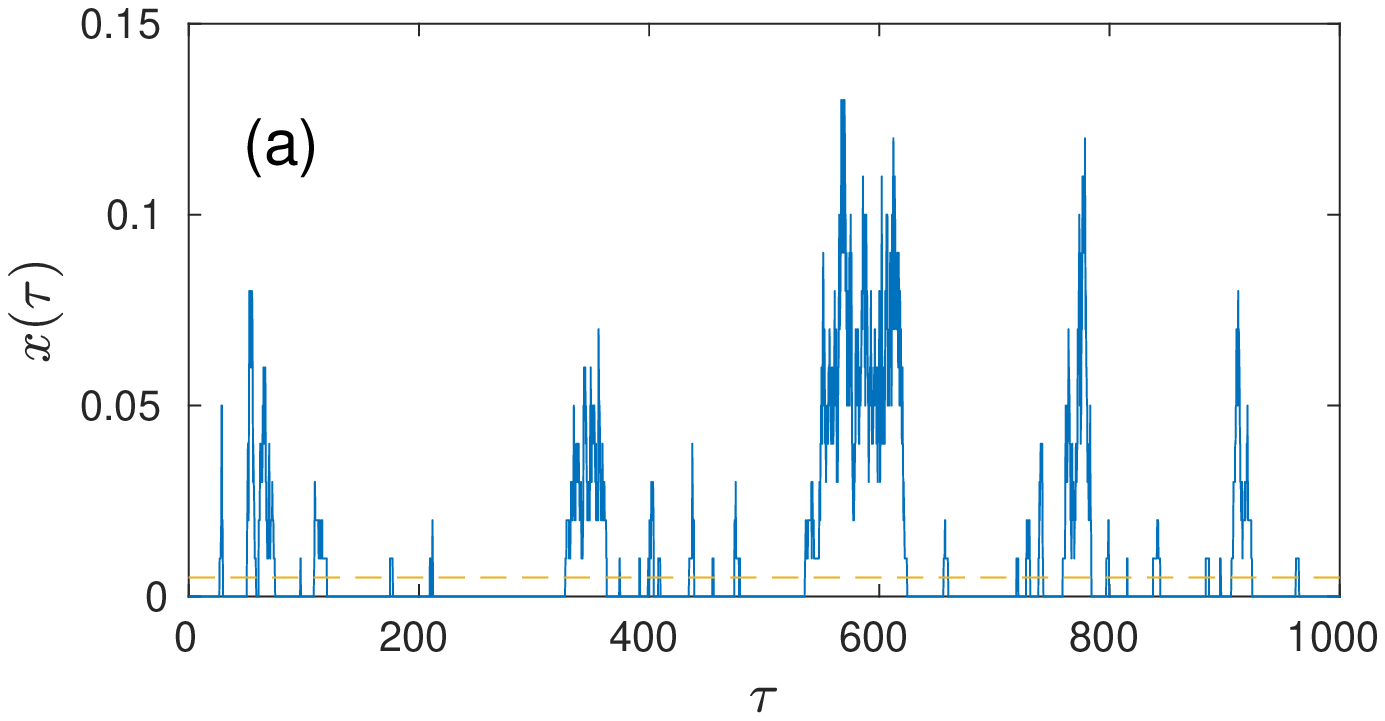}
	\includegraphics[width=0.49\textwidth]{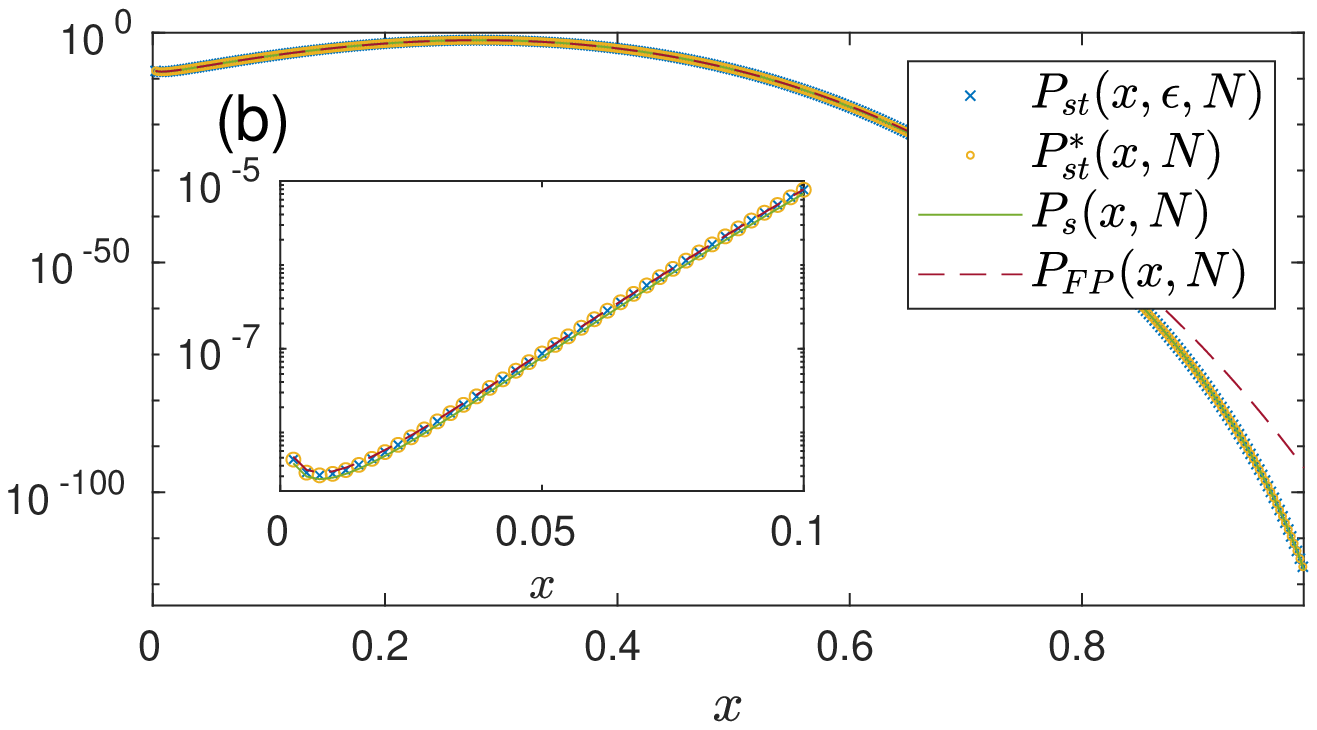}
	\includegraphics[width=0.49\textwidth]{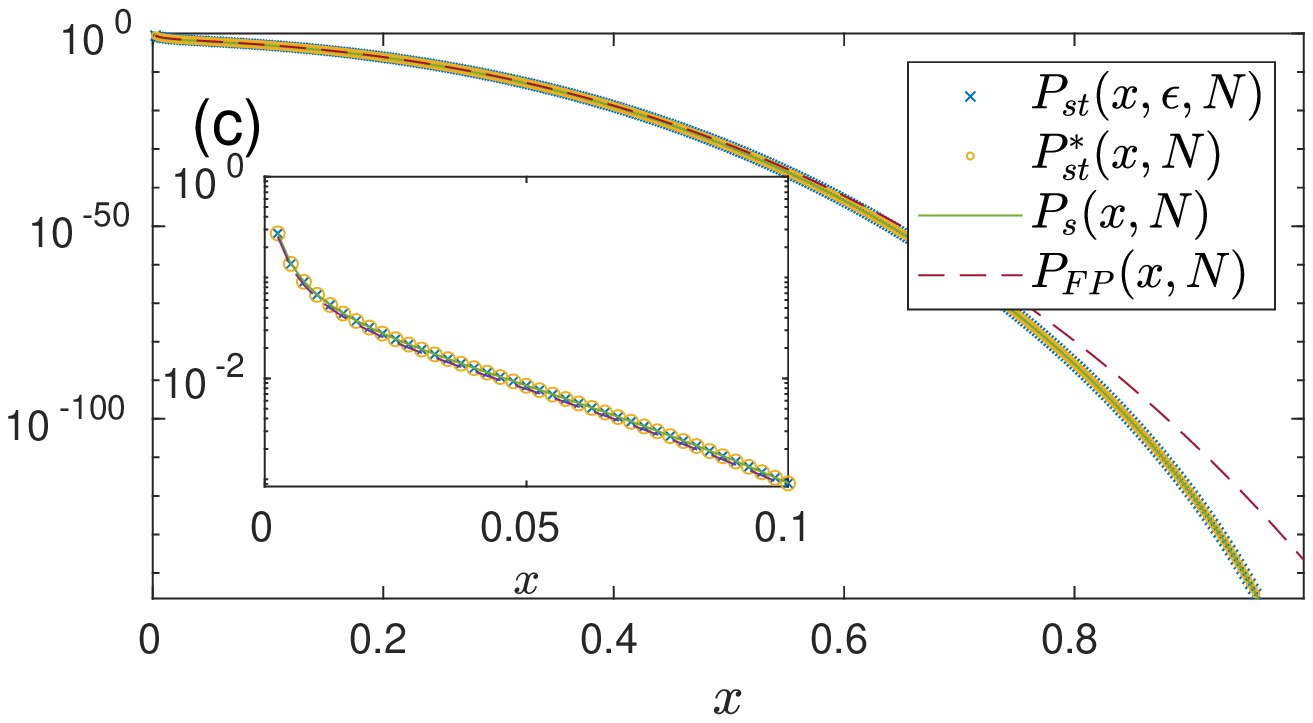}
	\includegraphics[width=0.49\textwidth]{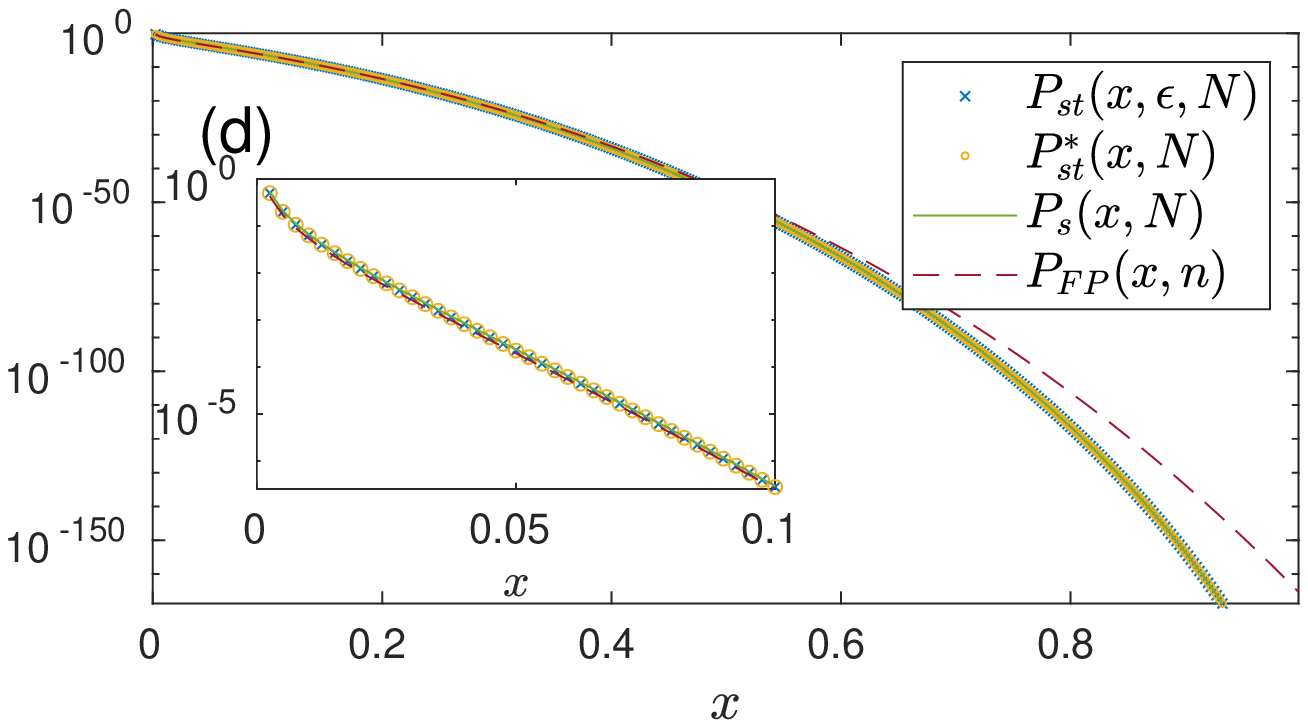}
	\caption{{\bf{Fluctuations and quasi-stationary state distributions.}}  Panel (a): the evolution of $x$ as a function of time $\tau=t/N$ with $g^-(x)=\mu x$ and $g^+(x,\epsilon)=\lambda (x+\epsilon)(1-x)$ ($\mu=.5$, $\lambda=.48$, $\epsilon=10^{-3}$, $N=100$).  In panels (b-d)  $P_{st}(x,\epsilon)$  has been obtained from Eq.\eqref{eq:prod} for $n\geq 0$, with $g^-(x)=\mu x$ and $g^+(x,\epsilon)=\lambda (x+\epsilon)(1-x)$. We plot only the solution for $n\geq 1 $ (i.e. $x\geq 1/N$) imposing the normalization on these sites ($\epsilon=10^{-10}$). The effective distribution $P^*_{st}(x,N)$ has been obtained for $x>1/N$ from Eq.\eqref{eq:prod2} setting $\epsilon=0$.  $P_S(x,N)$ and $P_{FP}(x,N)$ have given by Eq.\eqref{eq:stat8} and Eq.\eqref{eq:fokker3} at $\epsilon=0$ respectively. We fix $\mu=0.5$ and we describe the super-critical, critical ad sub-critical regimes by fixing $\lambda=0.7$, $\lambda=0.5$, $\lambda=0.4$ in panels (b), (c) and (d) respectively. Insets zoom in a small region around $x=0$, where the distributions diverge.} 
	\label{fig2}}
\end{figure}
As explicitly discussed in Appendix A,  $N F(x)$ is a sort of extensive free energy of the system, in analogy to what happens in equilibrium systems satisfying the detailed-balance condition. 

Before closing this preliminary section, let us also recall that
---as customarily done in the literature and done in detailed in Appendix B---
Eq.\eqref{eq:master} can be expanded in  power series of  $N$ (Kramers-Moyal expansion \cite{vK}), and its second-order truncation leads to a standard Fokker-Planck equation \cite{Gardiner,vK}.
As it has been already discussed, \cite{Garrido_1995,Doering_2005}, its associated stationary solution provides us with an accurate description of the exact stationary distribution Eq.\eqref{eq:prod} only around the maxima but fails to reproduce the statistics of the tails or rare events (see Appendices A and B).

\section{Quasi-Stationary distributions in the presence of an absorbing state}
\label{abs}

Let us explicitly consider the dynamics in the case  $\gamma^+(n_0)=0$, which implies, e.g., from
 Eq.\eqref{eq:prod},  that $n_0$  (that typically is the origin, i.e, $n_0=0$) is an absorbing state, which implies $P_{st}(n>0)=0$  and $P_{st}(0)=1$.  As a consequence, the only steady state distribution is a delta-Dirac at the origin.

{
In this context an interesting approach is obtained by introducing a small "spontaneous-creation" parameter $\epsilon$ that modifies the transition functions into $\gamma^+(n,\epsilon)$ and $\gamma^-(n,\epsilon)$ in such a way that $\gamma^+(0,\epsilon)>0$, $\lim_{\epsilon \to 0} \gamma^+(n,\epsilon)=\gamma^+(n)$ and $\lim_{\epsilon \to 0} \gamma^-(n,\epsilon)=\gamma^-(n)$
\cite{SIS1prod,SIS2prod}, so that the system has a non trivial stationary probability distribution $P_{st}(n,\epsilon)$. 
Let us remark  that in the limit of $\epsilon \to 0$,
$P_{st}(n,\epsilon)$ is expected to display interesting features, due to the presence of a critical transition to an absorbing state in the original model. In particular, for small enough $\epsilon$ and $n\geq 1$,
$P_{st}(n,\epsilon)$ depends on $\epsilon$ only through a global scaling factor,  i.e. 
$P_{st}(n,\epsilon)=h(\epsilon) P_{st}^*(n)$. The {\it quasi-stationary} normalized probability distribution  $P_{st}^*(n)$ 
\cite{Dickman1,Dickman2,Quasi-refs,Quasi} 
---i.e. the distribution conditioned to the fact that the system is active--- computed as }
\begin{equation}
P^*_{st}(n)= P^*_{st}(1) \prod_{j=2}^n \frac{\gamma^+(j-1)}{\gamma^-(j)}
\label{eq:prod2}
\end{equation}
with $\gamma^-(n)=\gamma^-(n,0)$ and $\gamma^+(n)=\gamma^+(n,0)$ (and where $P^*_{st}(1)$ needs to be fixed by imposing the normalization condition) {is independent of $\epsilon$.}
To make further progress, let us note that, in the limit of small $\epsilon$, $P_{st}(0,\epsilon)$ and the overall factor $h(\epsilon)$ are determined by Eq.\eqref{eq:stat} ($\gamma^+(0,\epsilon) P_{st}(0,\epsilon)=P_{st}(1,\epsilon)\gamma^-(1)=h(\epsilon)P^*_{st}(1)\gamma^-(1)$) together with the  normalization condition $P_{st}(0,\epsilon)+h(\epsilon)=1$.

As already discussed in \cite{SIS1prod,SIS2prod}, Eq.\eqref{eq:prod2} describes the stationary distribution of a model with a transition probability $\gamma^-(1)=0$ in the original SIS model at $\epsilon=0$,
which is an alternative prescription to avoid the system to be trapped in the absorbing state. 
Let us also remark that our approach is related to the method introduced by R. Dickman and collaborators {to describe \emph{quasi-stationary probability distributions}  in systems with absorbing states} \cite{Dickman1,Dickman2} (in the mathematical literature see, e.g., \cite{Quasi-refs,Quasi}).

By comparing Eq. \eqref{eq:prod2} with the procedure described in the previous section, we get that for $n\geq 1$, $P^*_{st}(n)$ should be well approximated for large enough $N$ by $P_S(x,N)$ given by Eq.\eqref{eq:stat8} with transition probabilities $\gamma^+(x)$ and $\gamma^{-}(x)$ evaluated at $\epsilon=0$. 
Since $n\geq 1$, a natural cut-off, i.e. $x\geq 1/N$, arises
in the continuous-limit case. In particular, let us remark that such a cut-off removes the divergence that is present for $x\to 0$ in the non-extensive term $(\gamma^-(x)\gamma^+(x))^{-1/2}$ in the distribution $P_S(x,N)$ in Eq.\eqref{eq:stat8} for systems with absorbing states.

Finally,  it is also possible to apply the continuous limit to the master equation and obtain a Fokker Planck equation, Eq.\eqref{eq:fokker}. As explained above, $P_{FP}(x,N)$ given by  Eq.\eqref{eq:fokker3} should provide us with a reliable estimate of the quasi-stationary probability distribution $P^*_{st}(n)$ around the maxima of the probability distribution.

Let us also remark that in a purely continuous approach with a Langevin Equation with multiplicative noise one obtains a continuous distribution with a non integrable singularity at the origin \cite{nature} similar to $P_S(x,N)$ or $P_{FP}(x,N)$. In the continuous case, however, there is no natural cutoff $1/N$, the probability distribution is not normalizable and the absorbing state $\delta(x)$ is the only stationary solution. In this perspective, our approach suggests a physical prescription to introduce a cut off in the diverging probability distribution of the continuous model, so that the regularized distribution describes the behavior of a discrete model where the collapse of the system in the absorbing state is forbidden by an arbitrary small escape probability.

To illustrate all this, in Figure \ref{fig2}  we present results from a computational simulation of   the standard SIS model, i.e. a paradigmatic example of a system with an absorbing state. 
The $\epsilon$ parameter is introduced by defining the transitions as $\gamma^-(x)=\mu x$, $\gamma^+(x,\epsilon)=\lambda (x+\epsilon)(1-x)$.
In panel (a) we plot a stochastic time series for the fraction of active sites $x$ as a function of time $\tau=t/N$ in the presence of a small $\epsilon=10^{-3}$, where $\lambda$ is set in the absorbing phase but close to criticality (as specified by the condition $\lambda_c=\mu$).  In panels (b-d) we consider the quasi-stationary distribution for different values of the parameters,  from the active phase (b), to the critical point (c), and
subcritical regime (d). In all cases, $P_{st}(x,\epsilon)$ has been obtained from the exact solution Eq.\eqref{eq:prod} at $\epsilon>0$, neglecting the probability to be in $x=0$ (i.e.  we consider only the evolution of $x$ during the excursion above the dashed line of panel (a)). 
$P^*_{st}(x)$ has been obtained from Eq.\eqref{eq:prod2} using $\gamma^+(x)$ with $\epsilon=0$. Observe that there is a perfect agreement between the two statistics, and the analytical expression $P_S(x,N)$ obtained for large $N$ for $1/N\leq x \leq 1$. Finally, as anticipated above, the Fokker-Planck approximation \eqref{eq:fokker2} and its relevant distribution \eqref{eq:fokker3} gives the correct behavior at the maximum but it fails in the large deviation regime, as expected. In the insets we zoom in a small region near $x=0$ in order to illustrate the divergence of the distribution and the cutoff at $z=1/N$

 Thus, in summary, we have illustrated that, in order to obtain \emph{bona fide} steady state distributions in systems with absorbing states it suffices to use a natural cutoff $1/N$ and assume that the state variable is confined to values equal or larger than it. This is, precisely, the strategy used in what follows.

\section{The q-SIS model}
\label{qsis} 

Let us consider a generalisation of the SIS model involving a higher-order interaction of $q$ sites, with transition probabilities given by 
 $\gamma^-(x)=\mu x^q$ and $\gamma^+(x)=\lambda x^{q} (1-x)$. For integer $q$, the model can be interpreted as a contact process with  transitions occurring only if a $q$-plet of infected sites are involved \cite{Carlon_2001,Park_2002,Odor}, i.e. 
\begin{eqnarray}
q {\rm I} \rightarrow  (q-1){\rm I}\ {\rm S}  & & \ \ {\rm with\ rate }\ \mu \nonumber \\
q {\rm I}\ {\rm S}  \rightarrow (q+1){\rm I} & & \ \  {\rm with\ rate }\ \lambda
\label{eq:ratesqSIS}
\end{eqnarray}
and the standard SIS process is recovered for $q=1$. 

The dynamics can be interpreted in terms of q-plet processes only for $q$ integer, however the  transitions $\gamma^+(x)$ and $\gamma^-(x)$ as a function of the total fraction $x$ are well defined for any $q>0$. In particular, for large $q>1$, the dynamics close to the absorbing state is slowed down, since the transition processes are less probable, while for small $q<1$ the dynamics speeds up.

\subsection{Mean-field dynamics}

The dynamics of the q-SIS model has been studied in the mean-field regime for $N\to \infty$ in \cite{Carlon_2001,Park_2002,Odor}. 
The deterministic mean-field equation controlling the density of active sites is simply \cite{vK,Gardiner,Odor-book}
\begin{equation}
\dot x(t) = -\mu x^q + \lambda x^q (1-x).
\label{eq:dynmf}
\end{equation}
For $\lambda>\mu$ the absorbing state is unstable and $x$ converges exponentially to the stable fixed point $x_0=1-\mu/\lambda$,  with a characteristic time $\tau$ that diverges at the criticality as $\tau\sim (\lambda-\mu)^{-q}$. For $\lambda<\mu$, the absorbing state $x=0$ is stable. In the standard SIS model ($q=1$), $x$ decays exponentially to zero as $\exp(-t/\tau)$ with the characteristic time $\tau=(\mu-\lambda)^{-1}$, that diverges at criticality $\lambda=\mu$. On the other hand, for $q>1$ a different behaviour is observed, namely at large times:
\begin{equation}
x(t) \sim t^{-1/(q-1)}.
\label{eq:dynmf3}
\end{equation}
Therefore, for higher-order processes, with $q>1$, a power-law decay emerges generically in the absorbing state, i.e., even away from the critical point. Finally, at criticality, $\lambda=\mu$, one has
\begin{equation}
x(t) \sim t^{-{1/q}}
\label{eq:dynmf5} 
\end{equation}
i.e. a power law is again observed, albeit with a slower decay.

These simple mean-field dynamical analyses reveal the crucial relevance of the parameter $q$ ---controlling the standard or higher-order nature of the process--- in determining dynamical scaling features \cite{Redner,Peliti,Cardy,Ben-Avraham,nature,Omar,Benitez,triplet,Odor-book}. 
Thus, in what follows, we wonder whether similar anomalous effects emerge in the stationary properties of this type of processes, for which we rely on the large-deviation approach.

{One could also consider the more general case where the birth process involves a different number $p$ of nodes (i.e. $\gamma^+(x)=\lambda x^p(1-x)$ with $p\not = q$). In this case for $p<q$ the active state is always stable and no transition can be observed for finite $\lambda$ and $\mu$. For $p>q$, the system becomes bistable and the transition between the active and the inactive phase is discontinuous. Therefore, the system does not present the critical behavior which typically characterizes second order continuous transitions. Therefore,  we focus on the non-trivial case $p=q$.}

\subsection{Finite-size scaling analyses}
\label{Fluct} 

Let us consider  the general analytic expression for the quasi-stationary probability distribution, $P_s(x,N)$,
as derived above to evaluate the average value and the relevant moments of $x$ as a function of the system size $N$ in the different phases. First of all, let us emphasize that, curiously enough, the effective free energy $F(x)=\log(\mu/\lambda)x +(1-x)\log(1-x)+x$ is independent of $q$. In other words, the exponent $q$, which drives the dynamics of $\langle x (t) \rangle$ in the infinite $N$ limit and, in particular, controls the time decay as shown by Eq.(\ref{eq:dynmf3}) and (\ref{eq:dynmf5}) appears only in the sub-leading non-extensive part of the quasi-stationary distribution: i.e., the parameter $q$ only affects the degree of the singularity at the origin:
\begin{equation}
P_S(x,N)=K \frac{e^{-N (\log(\mu/\lambda)x +(1-x)\log(1-x)+x)}}{x^q\sqrt{(1-x)}} 
\label{eq:stat_qSIS}
\end{equation}
with $1/N\leq x \leq 1$.
Let us remark that, in the whole physical regime $0\leq x\leq 1$, when the absorbing state is stable, i.e. $\lambda<\mu$, $F(x)$ has a minimum at $x=0$, while for $\lambda>\mu$ the absorbing state is dynamically unstable and $F(x)$ has a minimum in the stable fixed point of the mean-field evolution, $x_0=1-\mu/\lambda>0$. 

\vspace{0.5cm}
{\bf{Moments in the active phase.}}
In order to compute the moments of such distribution, Eq.(\eqref{eq:stat_qSIS}), 
let us first consider the active phase,  $\lambda>\mu$. Fixing $1/N<x*<x_0$, we can write
\begin{eqnarray}
\langle x^m \rangle&=& {\int_{1/N}^1 x^m P_S(x,N) dx} 
 \\
&=&{\int_{1/N}^{x^*} x^m P_S(x,N) dx}+
{\int_{x^*}^1 x^m P_S(x,N) dx} \nonumber
\label{eq:fluct2}
\end{eqnarray}
where the second integral can be easily estimated with a saddle-point approximation for large $N$ 
\begin{eqnarray}
&&\int_{x^*}^1 x^m P_S(x,N) dx \sim K \frac{x^m_0 e^{-NF(x_0)} }{x_0^q\sqrt{(1-x_0)}}  \int_{-\infty}^\infty e^{-N\frac{\lambda}{\mu}(x-x_0)^2}  dx \nonumber \\ &&= K \frac{\sqrt{\pi \mu}x^m_0 e^{-NF(x_0)} }{\sqrt{N\lambda}x_0^q\sqrt{(1-x_0)}}.
\label{eq:fluct3}
\end{eqnarray}
On the other hand, the first integral is instead determined by {the divergence  of  $P_S(x,N)$ at small values of} $x$:
\begin{equation}
\int_{1/N}^{x^*} x^m P_S(x,N) dx < \frac{ K e^{-NF(x^*)} }{\sqrt{(1-x^*)}} \int_{1/N}^{x^*} (x')^{m-q} dx' 
\label{eq:fluct4}
\end{equation}
and since $F(x^*)>F(x_0)$, Eq.\eqref{eq:fluct4} is exponentially suppressed in $N$ with respect to Eq.\eqref{eq:fluct3} and can therefore be neglected. Eq.\eqref{eq:fluct3} for $n=0$ fixes the normalization condition and fixes the value of $K$ as a function of $N$.
{We remark that for $x>0$ Eq.\eqref{eq:stat_qSIS} obeys a large deviation principle,} therefore,  in the limit of large $N$ we obtain that $\langle x^m \rangle \approx x_0^m$, since the saddle-point expansion around $x_0$ dominates the integral, {i.e. the probability accumulates at the mean value $x_0$}, while the divergence at $x=0$ with the natural cut-off $1/N$ can be discarded. 
One can also consider the fluctuations around the average value: since the saddle-point expansion in Eq.\eqref{eq:fluct3} displays Gaussian behaviour, fluctuations vanish for large $N$ as 
$\langle x^2 \rangle-\langle x \rangle^2 \approx 1/N$. Also in this case one can show that the effect of the divergence of $P_S(x,N)$ in $x=0$ is exponentially suppressed for large values of $N$. 

\vspace{0.5cm}
{\bf{Moments in the absorbing phase.}}
Let us now consider the case $\lambda<\mu$ for which the absorbing state is stable. Observe that, in this case,  $\dot F(0)=\log(\mu/\lambda)>0$, i.e. the derivative does not vanish at the origin ($x=0$). Let us choose an $x^*$ such that for $0<x<x^*$ one can approximate $F(x)\simeq \log(\mu/\lambda) x$ and $x^q \sqrt{1-x}\simeq x^q$. Then, it is possible to write:
\begin{equation}
\langle x^m \rangle \approx K {\int_{1/N}^{x^*}  \frac{x^m e^{-N x \log(\mu/\lambda)}}{x^q} dx}+ K
\int_{x^*}^1 \frac{x^m e^{-N F(x)}}{x\sqrt{1-x}} dx.
\label{eq:fluct5}
\end{equation}
The first integral can be solved setting $x'=Nx$, so that:
\begin{eqnarray}
&& K {\int_{1/N}^{x^*}  \frac{x^m e^{-N x \log(\mu/\lambda)}}{x^q} dx}
\simeq \nonumber \\ &&K N^{-m+q-1} \int_{1}^{\infty}  x'^{m-q} e^{- x' \log(\mu/\lambda)} dx'. 
\label{eq:fluct6}
\end{eqnarray}
Instead, for the second integral:
\begin{equation}
K \int_{x^*}^1 \frac{x^m e^{-N F(x)}}{x\sqrt{1-x}} dx
\leq K e^{-N F(x^*)} \int_{1/N}^1 \frac{x^m }{x\sqrt{1-x}} dx
\label{eq:fluct7}
\end{equation}
and since $F(x^*)>0$, Eq.\eqref{eq:fluct7} is exponentially suppressed for large $N$ with respect to Eq.\eqref{eq:fluct6} and therefore  it can be neglected. Eq.\eqref{eq:fluct5} for $n=0$ fixes the normalization constant $K$ and one has that the expectation value $\langle x^m \rangle$ vanishes with the system size as $\langle x^m \rangle \approx N^{-m}$ as expected if the absorbing state is stable. Moreover, the variance
 decays with $N$ as $\langle x^2 \rangle-\langle x \rangle^2\approx N^{-2}$. This means that, when the absorbing state is stable, the fluctuations in the system are much smaller than in the  Gaussian case, {which is just a consequence of the stable stationary state being an absorbing one.  To further illustrate this, observe that  considering the number $n$ of active sites instead of the fraction $x=n/N$ one readily obtains that ---independently of $q$--- all moments as well as the variances of the quasi-stationary probability distribution are finite (non-extensive) since they follow the distribution $e^{-n\log(\mu/\lambda)}/n^q$.}

$P^*_{st}(n)\approx e^{-n\log(\mu/\lambda)}/n^q$
\begin{figure}
	\centering{
	\includegraphics[width=0.49\textwidth]{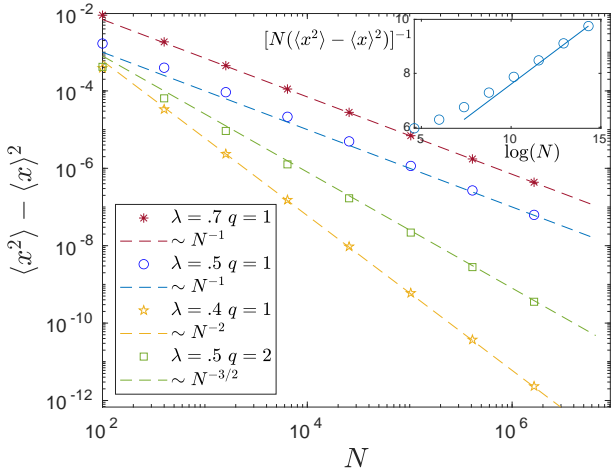}
	\caption{{\bf{Variance of the quasi-stationary distribution as a function of the system size $N$.}}
	Symbols refers to exact quantities evaluated numerically by means of the expression of Eq.\eqref{eq:prod} with $\gamma^+(x,\epsilon)=\lambda (x+\epsilon)(1-x)$ and $\gamma^-(x)=\mu x$ ($\mu=.5$ and $\epsilon=10^{-10}$). Dashed lines refer to the asymptotic expression: in the active phase ($\lambda=.7$, $\lambda>\mu$) we observe Gaussian fluctuations as $N^{-1}$; in the absorbing phase ($\lambda=.4$, $\lambda<\mu$) fluctuations decay as $N^{-2}$; at the critical point ($\lambda=\mu=0.5$) fluctuations are described by Eq.\eqref{eq:fluct10} (in the main plot we discard logarithmic corrections). In the inset we show that relevant logarithmic correction indeed occurs at the criticality $\lambda=\mu$ when $q=1$.
	}  	
	\label{fig3}}
\end{figure}

\vspace{0.5cm}
{\bf{Moments at the critical point.}}
 In the critical case,  $\mu=\lambda$,
 $F(x)=(1-x)\log(1-x)+x$ which can be approximated as $ 1/2 x^2$ at small $x$. One can introduce a small parameter $x^*$ such that the integral over $x>x^*$ can be neglected with respect to the integral over $x<x^*$. In this way, one is left with
\begin{eqnarray}
{\int_{1/N}^1 x^m P_S(x,N) dx} \simeq K {\int_{1/N}^{x^*}  \frac{x^m e^{-\frac{N x^2}{2}   }}{x^q} dx} \nonumber \\ \simeq N^{-(m-q+1)/2} K
{\int_{\frac{1}{N^{\frac{1}{2}}} }^{\infty}  \frac{x^m e^{- \frac{x^2}{2} }}{x^q} dx} \nonumber \\
\simeq \left\{
\begin{array}{ccc}
~~K(N)C_1(m) N^{-(m-q+1)/2} & {\rm if\ } & m>q-1\\ 
\\
K(N)C_2(m) \log(N) & {\rm if\ } & m=q-1\\
\\ ~K(N)C_3(m) N^{q-1-m} & {\rm if\ } & m<q-1.
\end{array}
\right.
\label{eq:fluct9}
\end{eqnarray}
Eq.\eqref{eq:fluct9} for $m=0$ provides the normalisation  condition for $P(x,N)$. One can first evaluate the decay to zero of the average number of active sites $\langle x \rangle$,  to obtain
the following set of expressions for different values of $q$:
\begin{equation}
\langle x \rangle  \sim \left\{
\begin{array}{ccc}
N^{-1/2} & {\rm if\ } & q<1\\
N^{-1/2} (\log(N))^{-1} & {\rm if\ } & q=1\\
N^{-(q+1)/2} & {\rm if\ } & 1<q<2\\
N^{-1} \log(N) & {\rm if\ } & q=2\\
N^{-1} & {\rm if\ } & q>2\\
\end{array}
\right.
\label{eq:fluct9b}
\end{equation}
The expression for the logarithmic corrections of $\langle x \rangle$ for the standard SIS model ($q=1$) had already been obtained ---directly from the exact formula \eqref{eq:prod2} for $P_{st}^*(n)$--- in \cite{SIS2prod}.

On the other hand, for the variance of the distribution one readily finds:
\begin{equation}
\langle x^2 \rangle - \langle x \rangle^2 \sim \left\{
\begin{array}{ccc}
N^{-1} & {\rm if\ } & q<1\\
(N \log(N))^{-1} & {\rm if\ } & q=1\\
N^{-\frac{q+1}{2}} & {\rm if\ } & 1<q<3\\
\log(N)N^{-2} & {\rm if\ } & q=3\\
N^{-2} & {\rm if\ } & q>3\\
\end{array}
\right.
\label{eq:fluct10}
\end{equation}

The asymptotic behaviors of fluctuations is illustrated in Figure \ref{fig3}. In particular, for $q<1$ the variance scale as in the Gaussian active phase, while for $q>3$ we recover the same scaling of the fluctuations as in the absorbing state (i.e. finite fluctuations of $n=x N$). 
The exact behaviour of the model is obtained from Eq.\eqref{eq:prod} with a small regularization parameter $\epsilon=10^{-10}$ (symbols). The exact results are compared, with the fluctuations of the active and the absorbing state for $\lambda>\mu$ and $\lambda<\mu$ respectively while they are compared with the asymptotic prediction of Eq.\eqref{eq:fluct10} in the critical regime $\lambda=\mu$. The plot reveals a very nice agreement between theory and numerics and elucidates, in particular, the presence of logarithmic corrections for $q=1$, as evinced in the inset.

\begin{figure}
	\centering{
	\includegraphics[width=0.49\textwidth]{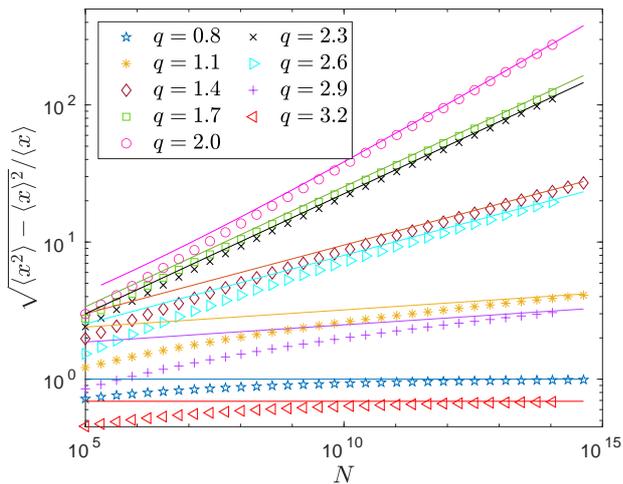}
	\caption{{\bf{Relative fluctuations, i.e. ratio ${\sqrt{\langle x^2 \rangle - \langle x \rangle^2}}/{\langle x \rangle}$ as a function of $N$ for several values of $q$.}} Symbols are obtained with exact evaluation of the stationary distribution by means of Eq.\eqref{eq:prod} with $\gamma^+(x,\epsilon)=\lambda (x+\epsilon)(1-x)$ and $\gamma^-(x)=\mu x$ we fix at the criticality $\mu=.5$ and $\lambda=0.5$; while for the small parameter we have $\epsilon=10^{-50}$. Lines correspond to the theoretical predictions for the different values of $q$ in Eq.\eqref{eq:fluct11}.  
	}
	\label{non_monotonic_q}}
	\end{figure}

Finally, it is illustrative  to compute the ratio between the variance and the mean, i.e., the relative weight of fluctuations:
\begin{equation}
\frac{\sqrt{\langle x^2 \rangle - \langle x \rangle^2}}{\langle x \rangle} \sim \left\{
\begin{array}{ccc}
C & {\rm if\ } & q<1\\
\\
\sqrt{\log(N)} & {\rm if\ } & q=1\\ \\
N^{{(q-1)/4}} & {\rm if\ } & 1<q<2\\ \\
N^{1/4}(\log(N))^{-1} & {\rm if\ } & q=2\\ \\
N^{(3-q)/4} & {\rm if\ } & 2<q<3\\ \\
\sqrt{\log(N)} & {\rm if\ } & q=3\\ \\
C & {\rm if\ } & q>3\\
\end{array}
\right.
\label{eq:fluct11}
\end{equation}
which exhibits a non-monotonic behavior as illustrated in Figure \ref{non_monotonic_q}:
 for $q<1$ and $q>3$ the ratio is constant (independent of $N$) while for $1\leq q \leq 3$ the ratio diverges with $N$, i.e.  fluctuations are much larger than the average at large $N$ even if both are vanishing. In particular, the ratio between the variance and mean grows the fastest with  $N$ for $q=2$. 
This last result emphasizes the crucial importance of the nature of the stochastic process, i.e. of $q$, in determining the nature of the critical fluctuating regime. In particular, relative fluctuations with respect to the mean are wild ---i.e. diverging with network size--- for higher-order interactions, around $q=2$.

\section{Conclusions}
\label{conc}

We have employed a large-deviation or WKB approach
to analyze the quasi-stationary distribution of general birth-death processes on fully-connected networks, exhibiting absorbing states. We have payed special attention to cases where more than one active node is required to generate further activity ---i.e. higher-order processes--- as exemplified by the $q$-SIS epidemic model. First of all, it has been shown (following existing results in the literature) that ---in order to regularize the problem and to avoid the system just falling asymptotically to the absorbing state--- one can either (i) introduce a small rate $\epsilon$ for the spontaneous generation of activity and then take the limit $\epsilon \rightarrow 0$ or (ii) constrain the system to have at least one active particle; these two approaches are equivalent and allow one to study a quasi-stationary distribution.

By using these combined techniques, we have been able to perform a finite-size analysis of all the moments of the quasi-stationary distribution of activity and elucidate a number of non-trivial features. First of all, in the
active phase the scaling is simply Gaussian. 
On the other hand, in the
absorbing phase, the variance of the quasi-stationary distribution scales with $N$ as $N^{-2}$ reflecting that fluctuations {are much more suppressed than} in the Gaussian case. Moreover, in this latter case, the distribution of the number of particles turns out to be an exponential.

Finally, as it is often the case, the situation is much more interesting at criticality, where we have found non-trivial expressions for the scaling of moments. In particular, we have shown that the variance-to-mean ratio diverges for $ N\rightarrow \infty$ for values of $q$ in the interval $[1,3]$ with the strongest divergence occurring at $q=2$. This anomalous scaling implies,  that fluctuations around the mean are much wilder when processes involving two-particles are at work. This also  emphasizes the  importance of the nature of the higher-order process, i.e. the value of $q$, in determining the nature of the critical fluctuating regime. 

As a general comment we want to explicitly remark once again that --owing to the presence of an absorbing state and its concomitant singularity at the origin--- the leading-order term
in a large-deviation approach \emph{does not suffice} to properly account for the steady state distribution: next-to-leading terms are crucial to obtain a sound description at criticality.

Let us also mention that the fact that the maximal variability is obtained for $q=2$, i.e. for the case in which "\emph{triplets}" are involved ---two sites creating activity plus one being activated--  is reminiscent of some recent findings e.g. (i) in theoretical ecology where triplets have been shown to stabilize ecological communities \cite{Jacopo} and  (ii) in neuroscience where triplet interactions (simplicial complexes) have been argued to be a minimal  crucial ingredient to rationalize neural data
\cite{Bassett}. We leave the exploration of the possible relation
between these observations for future work.

In a forthcoming work we plan to analyze the relation between the previous analysis of fluctuations in the quasi-stationary state, with the response to perturbations to the absorbing state, i.e.  with the statistics of avalanches at criticality.  We expect critical avalanches to be much more "volatile", i.e. to have a much larger variance, for the case $q=2$ exhibiting diverging variability,  but this needs to be confirmed by further numerical and analytical studies. These studies may have implications in the analyses of higher-order or "complex-contagion" processes of relevance e.g. in  actual epidemics, viral spreading, and models of opinion or belief propagation.
 
\appendix

\section{Mean-field dynamics with detailed balance}

Let us consider the special case of a system whose microscopic dynamics satisfies the detailed-balance condition. In this case there exists an equilibrium  distribution $P_E(x,N) \propto e^{-NV(x)}$. In particular, if $p^-(x)$ represents the probability to shift a variable from $1$ to $0$ and $p^+(x)$ is the probability of the reverse process (from $0$ to $1$), the detailed balance condition reads:
\begin{equation}
p^+(x)e^{-NV(v)}=p^-(x+1/N)e^{-NV(x+1/N)}.
\label{eq:bal1}
\end{equation}
Moreover, in this case one has
\begin{equation}
\gamma^-(x)=x p^-(x),\\\\\\\ \gamma^+(x)=(1-x)p^+(x) 
\label{eq:bal2}
\end{equation}
since
$x$ and $1-x$ represent the probabilities to select a variable in the state $1$ or $0$, respectively. 
Therefore, expanding Eq.\eqref{eq:bal1} for large $N$'s, one obtains:
\begin{eqnarray}
&&\log(p^+(x))-NV(v)=  \\ &&\log(p^-(x))+\frac{\dot{p}^-(x)}{{p}^-(x)}\frac{1}{N}-NV(x)-\dot{V}(x)-\frac{\ddot{V}(x)}{2}\frac{1}{N},  \nonumber
\label{eq:bal3}
\end{eqnarray}
and comparing terms of the same order in $1/N$:
\begin{eqnarray}
&& \dot{V}(x)=
\log(p^-(x))-\log(p^+(x)),
\nonumber \\ && \log(p^-(x))+\log(p^+(x))=C
\label{eq:bal4}
\end{eqnarray}
where $C$ is a constant. Plugging Eq.\eqref{eq:bal2} into Eq.\eqref{eq:stat8} and using Eq.\eqref{eq:bal4} one finally obtains:
\begin{equation}
P_S(x,N)\sim \frac{e^{\displaystyle{-N (V(x)+x\log(x)+(1-x)\log(1-x)))}}}{\sqrt{x(1-x)}}.
\label{eq:bal5}
\end{equation}
The detailed balance implies that a given configuration $\{\sigma_i\}$ has a probability $P_E(\{\sigma_i\})\sim e^{-NV(x(\{\sigma_i\}))}$.  Then, one readily has
\begin{equation}
P_E(x,N)\sim {e^{-N V(x)}}\frac{N!}{(xN)!((1-x)N)!} 
\label{eq:bal6}
\end{equation}
where the binomial factor represents
the numbers of states where  a fraction of nodes $x$ is in state $\sigma_i=+1$. Using the Stirling approximation for the factorials, one recovers $P_E(x,N)=P_S(x,N)$, which shows that the result in
Eq.\eqref{eq:stat8} leads to the correct prediction when the detailed-balance condition holds.

\section{A comparison with the standard Fokker-Planck equation approach}

\begin{figure}
	\centering{
	\includegraphics[width=0.49\textwidth]{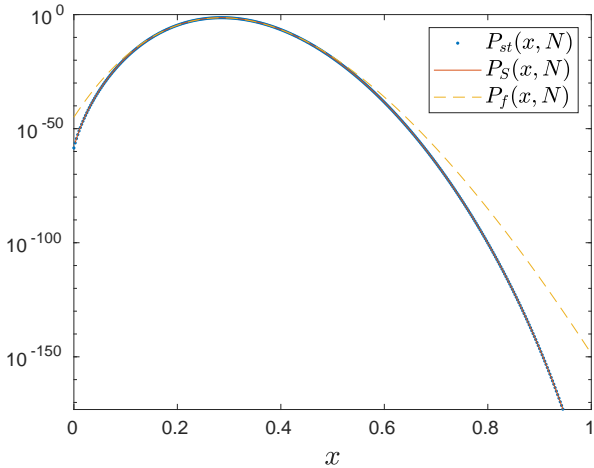}
	\caption{{\bf{ Stationary probability distribution for a mean-field dynamics.}} We take $\gamma^-(x)=0.5x$ and $\gamma^+(x)=0.2(1-x)$, i.e. at each step we chose randomly a variable; if it is in the state $+1$ the site is turned to $0$ with probability $0.5$ while, if it is in $0$, it is turned to $1$ with probability $0.2$ (N is  set to $N=400$). Dots represent the exact stationary distribution $P_{st}(x,N)$ computed using formula \eqref{eq:prod}. The continuous line represents the expansion for large $N$ large-deviation approach, $P_S(x,N)$ obtained in Eq.\eqref{eq:stat8} and the dashed line is the prediction obtained from the stationary solution $P_{FP}(x,N)$ of the Fokker-Planck equation.}  	
	\label{fig1}}
\end{figure}

The master equation \eqref{eq:master} can be rewritten as:
\begin{eqnarray}
 &&P(n,t+1)-P(n,t)  \nonumber\\ &=&\frac{1}{2}\{P(n+1,t)[\gamma^+(n+1)+\gamma^-(n+1)]
\nonumber\\ &&
+ P(n-1,t)[\gamma^+(n-1)+\gamma^-(n-1)]
\nonumber\\ &&
-2P(n,t)[\gamma^+(n)+\gamma^-(n)]\} \nonumber\\
& & +\frac{1}{2}\{P(n+1,t)[\gamma^-(n+1)-\gamma^+(n+1)]
\nonumber\\ &&
-P(n-1,t)[\gamma^-(n-1)-\gamma^+(n-1)]\}. 
\label{eq:fokker}
\end{eqnarray}
Introducing the fraction $x$ and taking the limit for large $N$ one obtains the usual Fokker-Planck equation:
\begin{eqnarray}
\frac{\partial P(x,\tau)}{\partial t} &=&
\frac{\partial}{\partial x}( (\gamma^-(x)-\gamma^+(x)) P(x,\tau) )
 \nonumber \\ &+&
\frac{1}{2N}
\frac{\partial^2 }{\partial x^2}(   (\gamma^-(x)+\gamma^+(x)) P(x,\tau))
\label{eq:fokker2}
\end{eqnarray}
where the time $\tau=t/N$, that can be considered a continuous variable, is measured in terms of $N$ microscopic steps.
Its associated stationary solution reads:
\begin{equation}
P_{FP}(x,N)\simeq K
\frac{\exp \left({\displaystyle{-N \int_c^x 2 \frac{\gamma^-(x')-\gamma^+(x')}{\gamma^-(x')+\gamma^+(x')} dx'}   } \right)}{\displaystyle{\gamma^-(x)+\gamma^+(x)}}   ~ 
\label{eq:fokker3}
\end{equation}
The stationary solution of Eq.\eqref{eq:fokker2} is different from the solution obtained from the master equation in the large $N$ limit in Eq.\eqref{eq:stat8}. In particular, in Figure \ref{fig1} we compare the exact distribution $P_{st}(x)$ given by Eq.\eqref{eq:prod}, the expansion for large $N$ in Eq.\eqref{eq:stat8} and the stationary solution of the Fokker Planck approach in Eq.\eqref{eq:fokker2}.
We observe that $P_{FP}(x,N)$ describes accurately the distribution around its maximum. However, rare events in the large deviation regime are given by Eq. \eqref{eq:stat8}. In this perspective, one can observe that around the maximum of the probability where the difference $\gamma^-(x)-\gamma^+(x)=\epsilon(x)$ is small, the expression in the exponential in Eq.\eqref{eq:fokker2} coincides with the exponential in Eq.\eqref{eq:stat8} up to the second order in the small parameter $\epsilon(x)$.

The difference between the stationary solution obtained via the Fokker Planck equation and the stationary solution in Eq.\eqref{eq:stat8}, which correctly describes the large deviation of the system, can be ascribed to a different expansion in the small parameter $1/N$. Let us expand directly in $1/N$ the master Equation \eqref{eq:master} in term of the fraction $x$. We get:
\begin{eqnarray}
&&P(x,t+1)-P(x,t)=-P(x,t)(\gamma^+(x)+\gamma^-(x))
\nonumber \\
&&+\sum_k \frac{1}{N^k k!} \frac{\partial^k P(x,t) \gamma^-(x,t) }{\partial x^k}+ 
\nonumber \\ &&
\sum_k \frac{(-1)^k}{N^k k!} \frac{\partial^k P(x,t) \gamma^+(x,t) }{\partial x^k}.
\label{eq:Fokkergen}
\end{eqnarray}
Clearly the terms $-P(x,t)(\gamma^+(x)+\gamma^-(x))$ exactly cancel out with the first term of the summations. Therefore, if we truncate the summation up to $k=2$ we exactly recover the Fokker Planck equation \eqref{eq:fokker2}. Let us now impose in Eq.\eqref{eq:Fokkergen} the stationary condition: $P(x,t+1)-P(x,t)=0$.
We now expand the stationary distribution according a large deviation formula  $P(x,t)=\exp(-NF(x)-g(x)-N^{-1} h(x)+\dots)$. If we plug this formula in \eqref{eq:Fokkergen} imposing that the first and the second terms in $1/N$ are vanishing, we get that $F(x)$ and $g(x)$ exactly satisfy conditions in Eq.\eqref{eq:stat7}. Therefore, in this way we obtain that the stationary distribution is given in the large $N$ limit by Eq.\eqref{eq:stat8}.

\vspace{1cm}
{\bf Acknowledgements:} MAM acknowledges the Spanish Ministry and Agencia Estatal de Investigaci{\'o}n (AEI) through Project of I+D+i Ref. PID2020-113681GB-I00, financed by 
MICIN/AEI/10.13039/501100011033 and FEDER "A way to make Europe", as well as the Consejer{\'\i}a de Conocimiento, Investigaci{\'o}n Universidad, Junta de Andaluc{\'\i}a and European Regional Development Fund, Project reference P20-00173 for financial support. We also thank  Roberto Corral and Pablo Hurtado  for useful comments and discussions.

\end{document}